\documentclass{aa}
\usepackage{graphicx}
\begin{document}
   \title{Iron abundances derived from RR~Lyrae light 
          curves and low-dispersion spectroscopy}

   \author{G. Kov\'acs\inst{}}


   \institute{
   Konkoly Observatory, P.O. Box 67, H-1525,
   Budapest, Hungary \\ \email {kovacs@konkoly.hu}
   \\
   }

   \date{January 21, 2005 / March 23, 2005 }

   \titlerunning {Comparison of iron abundances}
   \abstract{
With the aid of the All Sky Automated Survey (ASAS) database on the 
Galactic field, we compare the iron abundances of fundamental mode 
RR~Lyrae stars derived from the Fourier parameters with those obtained 
from low-dispersion spectroscopy. We show from a set of 79 stars, 
distinct from the original calibrating sample of the Fourier method 
and selected without quality control, that almost all discrepant 
estimates are the results of some defects or peculiarities either 
in the photometry or in the spectroscopy. Omitting objects deviating 
by more than $0.4$~dex, the remaining subsample of 64 stars yields 
Fourier abundances that fit the spectroscopic ones with 
$\sigma=0.20$~dex. Other, more stringent selection criteria and 
different Fourier decompositions lead to smaller subsamples and 
concomitant better agreement, down to $\sigma=0.16$~dex. Except 
perhaps for two variables among the 163 stars, comprised of the ASAS 
variables and those of the original calibrating set of the Fourier 
method, all discrepant values can be accounted for by observational 
noise and insufficient data coverage. We suggest that the agreement 
can be further improved when new, more accurate spectroscopic data 
become available for a test with the best photometric data. 
As a by-product of this analysis, we also compute revised periods 
and select Blazhko variables.        
 }

   \maketitle
%

%
%

\section{Introduction}
Determination of the chemical composition of RR~Lyrae stars is an 
important task because of the general astrophysical relevance of 
these objects. Unfortunately, RR~Lyrae stars have low brightness, 
and therefore most of them are accessible for more thorough 
spectroscopic investigations only by large telescopes. Furthermore, 
due to large variations in effective surface gravity and temperature 
during the pulsation, complicated hydrodynamical phenomena (moving 
ionization fronts and shocks) take place on time scales of an hour, 
which make the observations and analysis more difficult. For these 
reasons, high dispersion spectroscopic observations of RR~Lyrae stars 
are rather rare (see Butler et al. 1982; Clementini et al. 1994, 1995; 
Fernley \& Barnes 1996; Lambert et al. 1996; Sandstrom, Pilachowski 
\& Saha 2001), and aimed partially at the calibration of some variants 
of the spectral index method originally devised by Preston (1959). 
The overwhelming majority of the information available for us on the 
heavy abundance of RR~Lyrae stars comes from these spectral index 
methods. 

Although the acquisition of low-dispersion spectroscopic data is easier 
than that of high-dispersion data, with the increasing demand for  
studying more distant objects in dense stellar clusters and galaxies, 
the question arose of whether the more easily accessible light curves 
could be used for metallicity estimation. The fact that the period, 
amplitude and metallicity are correlated has been known since the 
pioneering paper of Preston (1959). However, the accuracy and the 
amount of the then-available data did not allow him to recognize the 
existence of a tighter correlation with another parameter associated 
with the shape of the light curve. By using accurate photometric and 
low-dispersion spectroscopic data, it was found that a simple linear 
formula, employing the period $P$ and the Fourier phase 
$\varphi_{31}$, correlates very well with the observed [Fe/H] and is 
able to predict the measured values with a standard deviation of 
$0.13$~dex (Jurcsik \& Kov\'acs 1996, hereafter JK96). 

Because of the accurate fit of the Fourier formula to the observed 
metallicities of the fundamental mode (RRab) field stars, it would 
be very attractive to use it on various large databases. Before one 
pursues such a task, it is useful to test the reliability of the 
method on various independent datasets with known metallicities. 
Several attempts have been made in this direction during the past 
several years. The first one was that of JK96 who compared globular 
cluster metallicities derived from the spectroscopy of giants with 
those computed by the Fourier method on the RRab stars of the 
respective clusters. The conclusion was that the overall agreement 
was satisfactory, but there was a hint that at the low-metallicity 
end the Fourier method gave somewhat higher values. Schwarzenberg-Czerny 
\& Kaluzny (1998) examined the correlation of the individual 
low-dispersion metallicities obtained by Butler, Dickens \& Epps 
(1978) for $\omega$~Cen. They concluded that the expected correlation 
was not seen. Subsequently, for the same cluster, Rey et al. (2000) 
found a correlation with the Fourier data when using metallicities 
derived from their {\it Caby} photometry. Borissova et al. (2004) 
compared the metallicities of a few stars in $\omega$~Cen derived 
from their modified $\Delta S$ method (adopted from Layden 1994, 
hereafter L94) with those obtained by Butler et al. (1987) and by 
Rey et al. (2000). They found that the former ones (used also by 
Schwarzenberg-Czerny \& Kaluzny 1998) were in better agreement 
with theirs than the latter ones. For M3, Jurcsik (2003) argued 
that the correlation of the direct spectroscopic observations of 
Sandstrom et al. (2001) with the ones derived by the Fourier method 
was significant, suggesting a real metallicity spread in this 
globular cluster. Based on a current sample of 22 RRab stars with 
low-dispersion spectroscopic abundances in the Large Magellanic 
Cloud (LMC), Gratton et al. (2004) concluded that, after selecting 
outliers, the derived standard deviation of $0.24$~dex between the 
Fourier and spectroscopic abundances were in agreement with the 
errors of the measurements.          

All these suggest that in spite of the various efforts to accumulate  
new spectroscopic abundances for RR~Lyrae stars, the accuracy and the 
amount of the data used in the above tests are insufficient to draw 
a firm conclusion on the reliability of the Fourier method. 
Unfortunately, there are no current large-scale low-dispersion surveys 
available and there is also a lack of similar undertakings on the 
high-dispersion side. There is also a need for extended multicolor 
photometry. Except for the current releases of photometric time 
series by the All Sky Automated Survey (ASAS) and the Robotic Optical 
Transient Search Experiment (ROTSE) projects (see Pojmanski 2002, 2003, 
Pojmanski \& Maciejewski 2004, Wozniak et al. 2004) and perhaps for 
the Behlen Observatory survey (Schmidt \& Seth 1996) and that of 
Layden (1997), no significant efforts have been made to extend the 
photometric database for Galactic field RR~Lyrae stars since the 
formula of JK96 was derived. 

For a useful test of the Fourier formula, we need reliable light curves 
observed in Johnson $V$ band, with corresponding [Fe/H] values either 
from L94 or from Suntzeff, Kraft \& Kinman (1994, hereafter S94). 
(We recall that it was shown by JK96 that the two databases are 
statistically equivalent.) Various reasons (no $V$ band observations 
by ROTSE, few data points per star in the Behlen survey, too few 
variables with proper light curves and overlaps with the ASAS database 
in the survey of Layden 1997) led us to utilize only the database of 
the ASAS survey. Although we think that this database, combined with 
the metallicities of L94, constitutes the most stringent current test, 
it is important to keep in mind its limitations. Most importantly, 
many of the variables are rather faint and therefore, due to the small 
aperture telescopes employed by the ASAS project, they exhibit relatively 
large scatter (compensated partially by the large number of data points 
per star). In addition, no time-dependent color term was employed in 
the transformation to the standard magnitude system, therefore, small 
distortions of the published light curves are expected. The fainter 
objects used in our tests are also more likely to show larger errors 
in their measured metallicities. Nevertheless, as it is shown later, 
the present test supports the applicability of the Fourier method in 
the determination of the iron abundances of RRab stars.

%
%

\section{Light curves and metallicities}
%
%
%
%
%
%
%
%
\begin{table*}[t]                                                     
\caption[]{ASAS variables common with those in the compilation of JK96}                                                    
\begin{flushleft}                                                     
\begin{tabular}{rcrcrcrcrcrc}                                         
\hline                                                                
Variable & Period [d] & [Fe/H] & Rem &                                
Variable & Period [d] & [Fe/H] & Rem &                                
Variable & Period [d] & [Fe/H] & Rem\\                                
\hline
\object{WY Ant}     & 0.574341 & $-1.39$ & 1D  & 
\object{SV Hya}     & 0.478548 & $-1.43$ & BL  & 
\object{TY Pav}     & 0.710382 & $-2.01$ & $-$\\
\object{V Cae}      & 0.570914 & $-1.71$ & $-$ & 
\object{DH Hya}     & 0.489002 & $-1.28$ & $-$ & 
\object{U Pic}      & 0.440375 & $-0.50$ & 1D \\
\object{IU Car}     & 0.737059 & $-1.57$ & LD  & 
\object{FY Hya}     & 0.636650 & $-2.03$ & $-$ & 
\object{RY Psc}     & 0.529746 & $-1.18$ & BL \\
\object{V499 Cen}   & 0.521212 & $-1.29$ & 1D  & 
\object{SS Leo}     & 0.626337 & $-1.56$ & $-$ & 
\object{BB Pup}     & 0.480549 & $-0.35$ & LD \\
\object{RR Cet}     & 0.553028 & $-1.29$ & $-$ & 
\object{TV Leo}     & 0.672860 & $-1.86$ & $-$ & 
\object{AV Ser}     & 0.487563 & $-0.95$ & $-$\\
\object{RX Cet}     & 0.573727 & $-1.20$ & BL  & 
\object{U Lep}      & 0.581479 & $-1.67$ & $-$ & 
\object{RW Tra}     & 0.374039 & $ 0.27$ & $-$\\
\object{RZ Cet}     & 0.510600 & $-1.24$ & BL  & 
\object{TV Lib}     & 0.269623 & $-0.17$ & $-$ & 
\object{W Tuc}      & 0.642236 & $-1.37$ & $-$\\
\object{W Crt}      & 0.412013 & $-0.45$ & $-$ & 
\object{VY Lib}     & 0.533940 & $-1.06$ & $-$ & 
\object{UU Vir}     & 0.475608 & $-0.60$ & $-$\\
\object{RX Eri}     & 0.587247 & $-1.07$ & $-$ & 
\object{RV Oct}     & 0.571170 & $-1.08$ & 1D  & 
\object{AT Vir}     & 0.525777 & $-1.54$ & $-$\\
\object{BB Eri}     & 0.569910 & $-1.23$ & LD  & 
\object{ST Oph}     & 0.450356 & $-1.02$ & $-$ &
$-$                 & $-$      & $-$     & $-$\\
\object{SX For}     & 0.605342 & $-1.35$ & $-$ & 
\object{V445 Oph}   & 0.397023 & $ 0.01$ & $-$ &  
$-$                 & $-$      & $-$     & $-$\\
\hline                                                                
\end{tabular}                                                         
\end{flushleft}                                                       
{\footnotesize\underline {Notes:} 
$\rm [Fe/H]$ : spectroscopic iron abundance as given in JK96; 
BL : Blazhko variable; 
1D : light curve is affected by an instrumental effect of $n$d$^{-1}$ 
frequency ($n=1, 2, ...$);                                                
LD : long-term drift or periodic variation in the average light level
}                                          
\end{table*}
%

%
%
%
%
%
%
\begin{table*}[t]                                                     
\caption[]{ASAS variables separate from those in the compilation of JK96}                                                    
\begin{flushleft}                                                     
\begin{tabular}{rcrcrcrcrcrc}                                         
\hline                                                                
Variable & Period [d] & [Fe/H] & Rem &                                
Variable & Period [d] & [Fe/H] & Rem &                                
Variable & Period [d] & [Fe/H] & Rem\\                                
\hline
\object{TY Aps    } & 0.501699 & $-0.96$ & 1D  & 
\object{SW Dor    } & 0.525338 & $-1.33$ & $-$ & 
\object{UV Oct    } & 0.542582 & $-1.34$ & BL \\
\object{UY Aps    } & 0.482513 & $-0.45$ & $-$ & 
\object{SX Dor    } & 0.631449 & $-1.71$ & $-$ & 
\object{AR Oct    } & 0.394029 & $-0.29$ & LD \\
\object{VX Aps    } & 0.484665 & $-1.18$ & BL  & 
\object{VW Dor    } & 0.570569 & $-0.99$ & BL  & 
\object{V413 Oph  } & 0.449001 & $-0.76$ & $-$\\
\object{XZ Aps    } & 0.587268 & $-1.30$ & 1D  & 
\object{XY Eri    } & 0.554272 & $-1.79$ & BL  & 
\object{V1023 Oph } & 0.623930 & $-0.96$ & 1D \\
\object{AR Aps    } & 0.610717 & $-1.49$ & $-$ & 
\object{BK Eri    } & 0.548165 & $-1.37$ & $-$ & 
\object{WY Pav    } & 0.588572 & $-0.74$ & 1D \\
\object{BS Aps    } & 0.582561 & $-1.33$ & 1D  & 
\object{AC Eri    } & 0.482083 & $-1.06$ & $-$ & 
\object{TZ Phe    } & 0.615603 & $-1.04$ & $-$\\
\object{CK Aps    } & 0.623631 & $-1.23$ & BL  & 
\object{AL Eri    } & 0.656941 & $-1.61$ & BL  & 
\object{HH Pup    } & 0.390746 & $-0.46$ & $-$\\
\object{DD Aps    } & 0.648108 & $-0.81$ & $-$ & 
\object{RX For    } & 0.597317 & $-1.01$ & BL  & 
\object{X Ret     } & 0.492016 & $-1.06$ & BL \\
\object{DI Aps    } & 0.519180 & $-1.33$ & $-$ & 
\object{SS For    } & 0.495434 & $-1.09$ & BL  & 
\object{V765 Sco  } & 0.463662 & $-0.93$ & $-$\\
\object{EL Aps    } & 0.579722 & $-1.29$ & $-$ & 
\object{SW For    } & 0.803760 & $-1.67$ & $-$ & 
\object{RU Scl    } & 0.493356 & $-1.00$ & $-$\\
\object{ER Aps    } & 0.758417 & $-1.13$ & $-$ & 
\object{UU Hor    } & 0.643691 & $-1.52$ & $-$ & 
\object{VX Scl    } & 0.637047 & $-1.95$ & $-$\\
\object{EX Aps    } & 0.471801 & $-0.58$ & 1D  & 
\object{SZ Hya    } & 0.537228 & $-1.48$ & BL  & 
\object{WY Scl    } & 0.463685 & $-1.25$ & $-$\\
\object{LU Aps    } & 0.755038 & $-1.77$ & $-$ & 
\object{WZ Hya    } & 0.537718 & $-1.04$ & $-$ & 
\object{AE Scl    } & 0.550113 & $-1.61$ & $-$\\
\object{U Cae     } & 0.419782 & $-0.86$ & $-$ & 
\object{XX Hya    } & 0.507752 & $-1.07$ & $-$ & 
\object{RV Sex    } & 0.503417 & $-0.85$ & BL \\
\object{KS Cen    } & 0.397445 & $-0.57$ & $-$ & 
\object{DG Hya    } & 0.754242 & $-1.16$ & $-$ & 
\object{AE Tuc    } & 0.414529 & $-0.57$ & 1D \\
\object{V671 Cen  } & 0.779058 & $-1.14$ & $-$ & 
\object{ET Hya    } & 0.685521 & $-1.42$ & $-$ & 
\object{AG Tuc    } & 0.602584 & $-1.67$ & $-$\\
\object{V674 Cen  } & 0.493935 & $-1.26$ & 1D  & 
\object{FX Hya    } & 0.417338 & $-0.83$ & $-$ & 
\object{BK Tuc    } & 0.550068 & $-1.54$ & $-$\\
\object{UU Cet    } & 0.606073 & $-1.74$ & $-$ & 
\object{GS Hya    } & 0.522831 & $-1.48$ & BL  & 
\object{FS Vel    } & 0.475754 & $-0.92$ & $-$\\
\object{RU Cet    } & 0.586284 & $-1.33$ & BL  & 
\object{IV Hya    } & 0.532251 & $-1.08$ & $-$ & 
\object{ST Vir    } & 0.410813 & $-0.64$ & $-$\\
\object{RV Cet    } & 0.623410 & $-1.06$ & BL  & 
\object{TW Hyi    } & 0.675383 & $-1.39$ & $-$ & 
\object{UV Vir    } & 0.587084 & $-0.94$ & $-$\\
\object{RV Cha    } & 0.701400 & $-1.39$ & $-$ & 
\object{XX Lib    } & 0.698452 & $-1.21$ & 1D  & 
\object{WW Vir    } & 0.651559 & $-1.84$ & $-$\\
\object{RT Col    } & 0.536580 & $-1.17$ & $-$ & 
\object{T Men     } & 0.409820 & $-0.61$ & $-$ & 
\object{AM Vir    } & 0.615083 & $-1.19$ & BL \\
\object{RW Col    } & 0.545615 & $-0.79$ & $-$ & 
\object{RX Men    } & 0.456290 & $-1.27$ & 1D  & 
\object{AS Vir    } & 0.553436 & $-1.23$ & $-$\\
\object{RX Col    } & 0.593766 & $-1.43$ & BL  & 
\object{EM Mus    } & 0.467286 & $-1.28$ & $-$ & 
\object{BQ Vir    } & 0.637029 & $-1.06$ & BL \\
\object{RY Col    } & 0.478840 & $-0.86$ & BL  & 
\object{Y Oct     } & 0.646571 & $-1.26$ & 1D  & 
\object{SV Vol    } & 0.609919 & $-1.89$ & BL \\
\object{X Crt     } & 0.732835 & $-1.48$ & BL  & 
\object{RS Oct    } & 0.457999 & $-1.33$ & BL  & 
$-$                 & $-$      & $-$     & $-$ \\
\object{RT Dor    } & 0.482834 & $-1.26$ & $-$ & 
\object{SS Oct    } & 0.621850 & $-1.33$ & BL  & 
$-$                 & $-$      & $-$     & $-$ \\
\hline                                                                
\end{tabular}                                                         
\end{flushleft}                                                       
{\footnotesize\underline {Notes:} 
$\rm [Fe/H]$ : Spectroscopic iron abundance as given in 
Layden (1994) or in Suntzeff et al. (1994);                                 
BL, 1D, LD : as in Table~1}                                          
\end{table*}                                                                    

In 2004 the ASAS project completed the variability survey of the 
$0^{\rm h}-18^{\rm h}$ segment of the southern hemisphere. Among the 
many other variables, they classified 723 RRab stars. All light curves 
contain some 100--700 data points. We searched the ASAS web 
site\footnote{http://www.astrouw.edu.pl/$\sim$gp/asas/} for RRab 
variables found also in the [Fe/H] catalogs of L94 or S94. We found 
110 variables satisfying this criterion. The data cover a period of 
3.7~y, giving an ample timebase for the analysis of the stability of 
the light variation. For each variable, the database contains five 
time series, corresponding to five different aperture sizes used in 
the photometry. All variables have been subjected to an analysis 
consisting of the following steps.
\begin{itemize}
\item
Time series $V(t)$ were generated through the weighted average of the 
magnitude values $V_i(t)$ given for the five apertures. By using the 
assigned standard deviations $\sigma_i(t)$, we have 
%
%
\begin{eqnarray}
V(t) & = & \sum_{i=1}^5 V_i(t)\sigma^{-2}_i(t)
\over {\sum_{i=1}^5 \sigma^{-2}_i(t)} \nonumber
\end{eqnarray}
\item
Frequency analysis of $V(t)$ was performed in the [0,3]~d$^{-1}$ 
interval, searched for the best period with a non-linear least 
squares method and thereafter prewhitened in order to find possible 
additional components. We employed the time series analysis program 
package {\sc mufran} (Koll\'ath, 1990)
\item
In order to avoid overfitting noisy light curves and underfitting 
those with low noise, after manual experiments, we use the following 
condition on the order of the Fourier fit. For each aperture, the 
light curve was fitted by a Fourier sum of order $m$, computed from 
the following condition
%
%
\[ m = \left\{ \begin{array}{ll}
4              & \mbox{if $m^{\star}<4$} \\
$$m^{\star}$$  & \mbox{if $4\leq m^{\star}\leq 10$} \\
10             & \mbox{if $m^{\star}>10$}
               \end{array}
       \right. \]
Here the `empirical' order $m^{\star}$ is given by  
$m^{\star}=INT({\rm SNR}/10)$, where the symbol $INT$ stands for 
the integer part and the the signal-to-noise ratio. The latter 
quantity is computed by 
%
%
\begin{eqnarray}
{\rm SNR} & = & A_1\sqrt{n}/\sigma \nonumber
\end{eqnarray}
where $A_1$ is the Fourier amplitude at the main frequency component, 
$n$ is the number of data points and $\sigma$ is the standard deviation 
of the residuals between the data and the fit.
\item
Outliers were discarded at the $3\sigma$ level and the Fourier fit 
was repeated until no outliers were found. 
\item
Finally we stored the light curve corresponding to that aperture size 
which yielded the highest SNR.
\end{itemize} 
%

%
%
%
   \begin{figure*}[t]
   \centering
   \includegraphics[width=130mm]{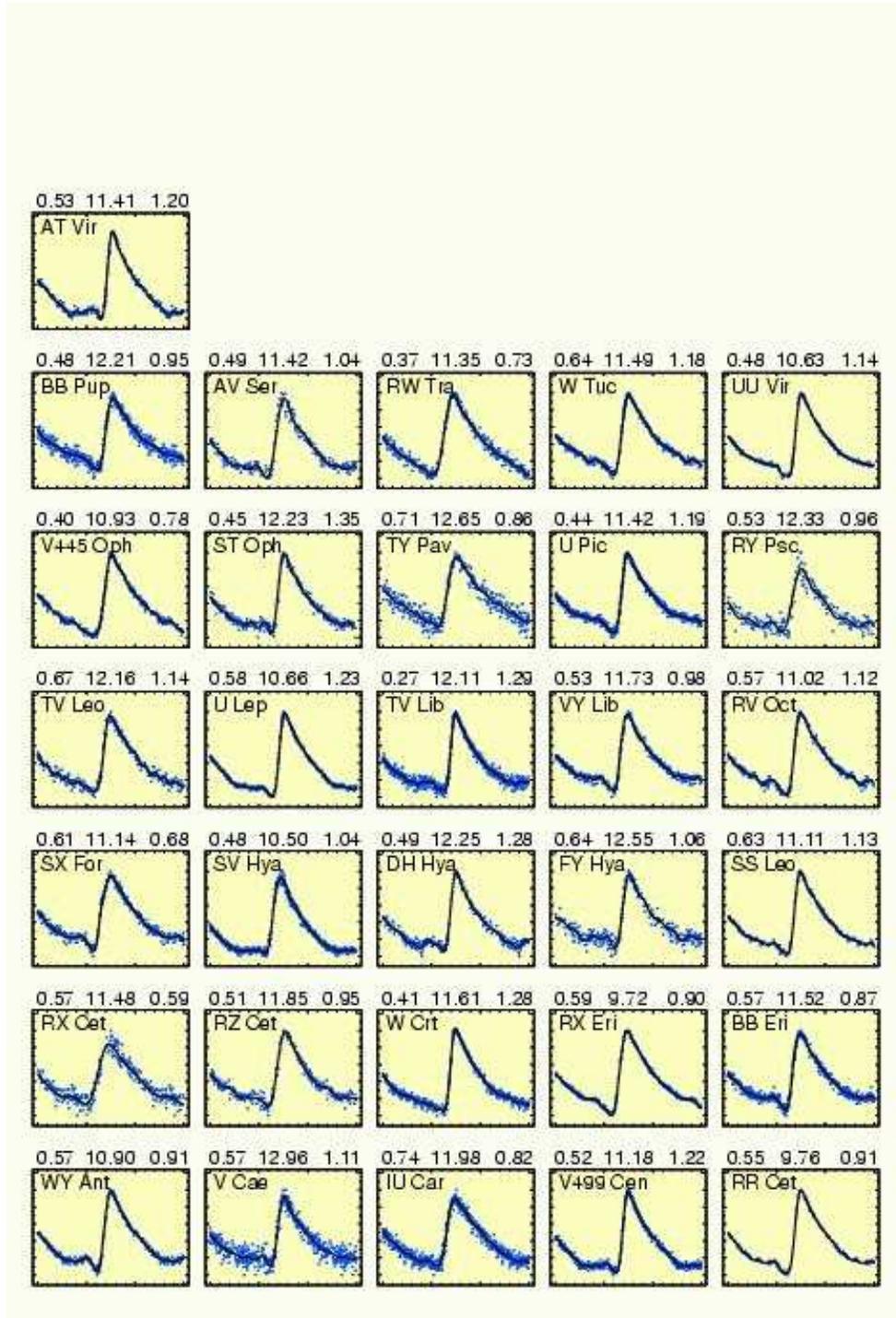}
      \caption{Folded light curves from the ASAS database. All variables 
               enter also in the compilation of JK96. {\em Headers:} 
	       period [d], average brightness in $V$ and total amplitude 
	       [mag] of the fitted light curve (continuous line). The 
	       orders of the Fourier fits were determined from the SNR 
	       of the light curves as described in the text. All light 
	       curves were obtained in Johnson $V$ color. With centered 
	       peak brightness, the light curves are plotted in the 
	       $1.5$ phase interval.} 
         \label{fig1}
   \end{figure*}
%
%

%
%
%
   \begin{figure*}[t]
   \centering
   \includegraphics[width=130mm]{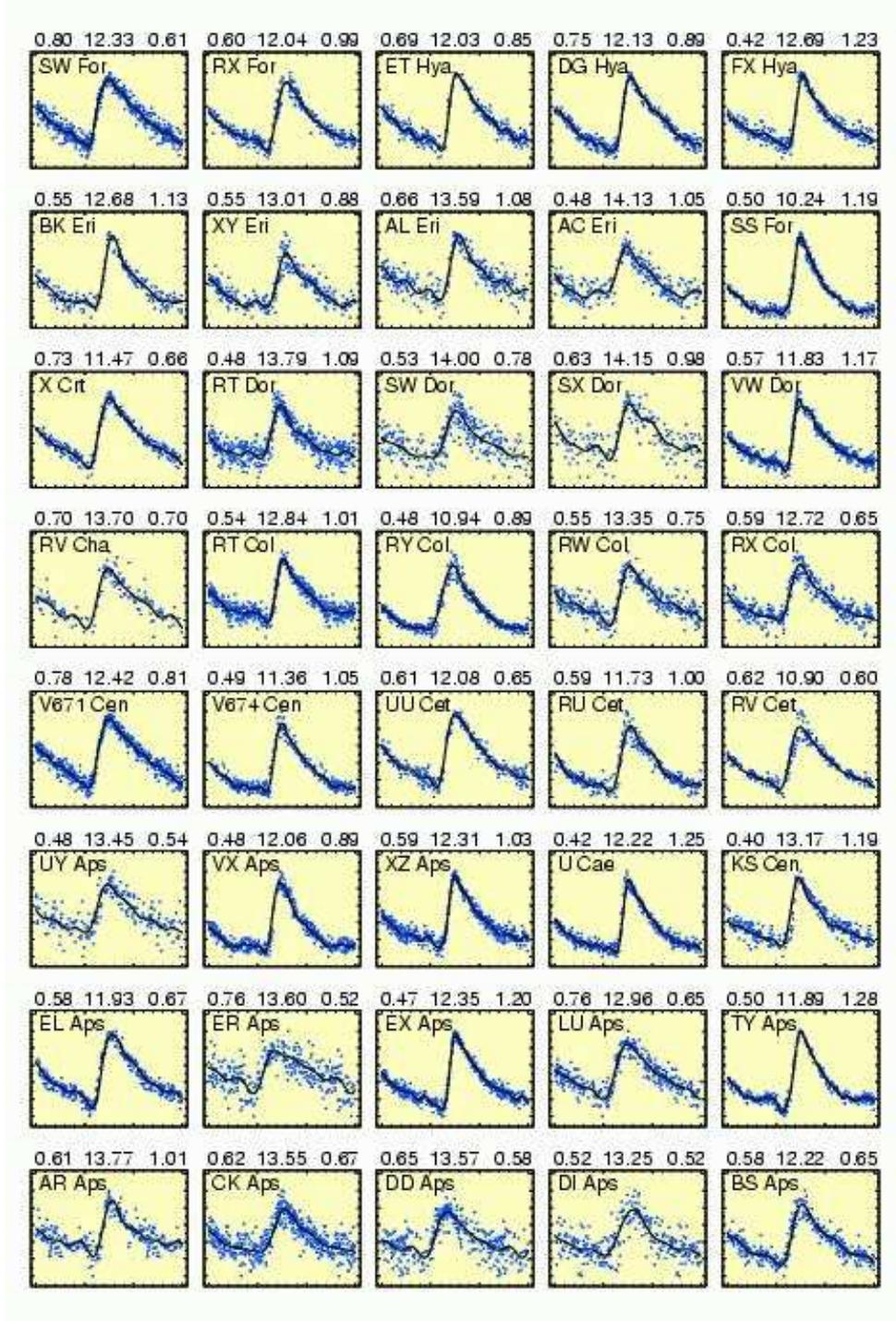}
      \caption{Folded light curves from the ASAS database. None of the 
               variables enter in the compilation of JK96 but all 
	       have low-dispersion spectroscopic metallicities either 
	       from Layden (1994) or from Suntzeff et al. (1994). 
	       Notation is the same as in Fig.~\ref{fig1}.} 
         \label{fig2}
   \end{figure*}
%
%

%
%
%
   \begin{figure*}[t]
   \centering
   \includegraphics[width=130mm]{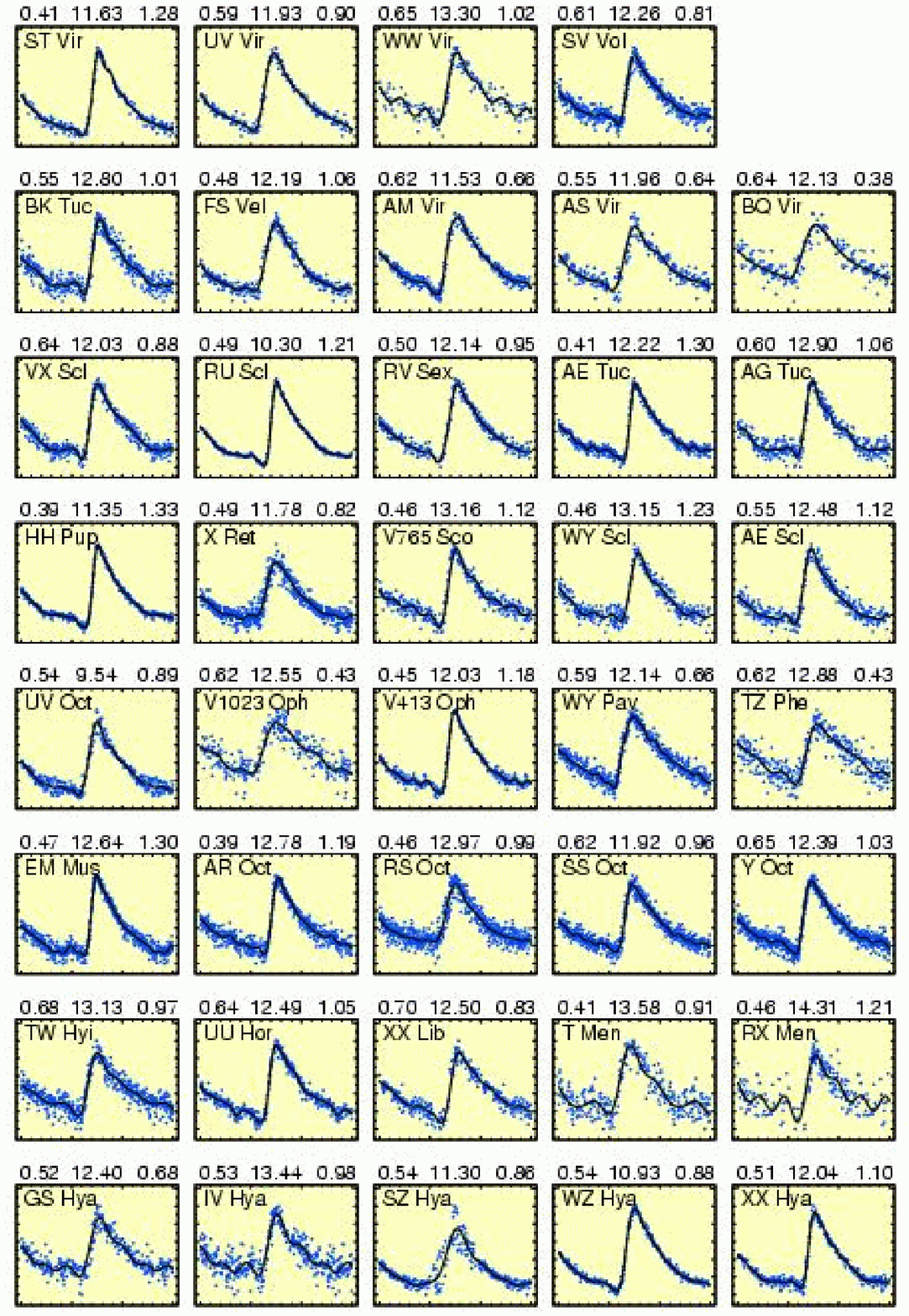}
      \caption{Folded light curves from the ASAS database. Data selection 
               and notation are the same as in Fig.~\ref{fig2}.} 
         \label{fig3}
   \end{figure*}

The set of 110 variables is separated into two groups. The first 
group (set \#1, with 31 stars) is common with a subset of the 
variables used by JK96. This set is used primarily to compare the 
ASAS photometry with the one of JK96, that is based on traditional 
photometric observations of various origins. The second group 
(set \#2, with 79 stars) has no common members with the dataset of 
JK96. This set is used to check the predictive power of the Fourier 
formula for [Fe/H]. Variable names, derived periods (based on the 
ASAS database), iron abundances and comments on the light variations 
of the individual variables are listed in Tables~1 and 2 for sets 
\#1 and \#2, respectively. Folded light curves with the corresponding 
Fourier fits are displayed in Figs.~\ref{fig1}, \ref{fig2} and \ref{fig3}.

Iron abundances in Table~1 are the same as given in JK96. We recall 
that except for two stars among the 84 variables entering in the 
compilation of JK96, all abundances were computed from the weighted 
and zero-point-shifted averages of S94 and L94. The metallicity 
scale of Jurcsik (1995) is used that corresponds to the scale of 
the high-dispersion spectroscopic studies and is similar to those 
of Clementini et al. (1995) and Lambert et al. (1996), with these 
latter calibrations being more metal-poor by 0.1--0.15~dex. The 
abundances in Table~2 were obtained almost exclusively from L94. 
For UU~Cet and AL~Eri we utilized the data of S94.

The frequency analysis (combined with prewhitening) allowed us to 
identify a number of Blazhko (BL) variables, together with those 
showing integer d$^{-1}$ periodicities in the prewhitened spectra 
(1D stars) and with those displaying long-periodic variations in 
their average light levels (LD stars). Due to aliasing, in some 
cases ambiguity might exist between the 1D and LD classifications. 
Except perhaps for a few cases, these variations are thought to be 
related to some anomaly in the method employed in the data reduction 
pipeline. These types of systematics are common in almost all 
photometric databases, but especially in the ones originating from 
wide field photometry (see Kov\'acs, Bakos \& Noyes 2005 and 
references therein).   

Classification of the BL stars was based on the occurrence of nearby 
component(s) at the fundamental pulsation frequency. We found 26 BL 
variables from the 110 RRab stars. Most of these are new discoveries, 
some of them (e.g., \object{RU~Cet}, \object{SV~Vol}) show weak, 
whereas others (e.g., \object{RX~Cet}, \object{SZ~Hya}) show strong 
Blazhko effects. The derived incidence rate of 24\% is in agreement 
with the one found by Moskalik \& Poretti (2003) from the analysis 
of 150 RRab stars in the Galactic bulge and also in line with the 
earlier quoted rates for RRab stars in general (Szeidl 1988). The 
above rate for the LMC is only 12\% (Alcock et al. 2003), and it is 
tempting to attribute this lower value to the lower average metallicity 
of the LMC. Nevertheless, by checking the entries in our tables, it 
seems to us that one cannot draw such a simple conclusion, since 
both low- and high-metallicity stars can be found among the Galactic 
field BL variables. 

By inspecting the folded light curves, we see that most of them are 
of good or fine quality, in spite of the expected large scatter at 
these faint magnitudes for photometric data obtained by a small-aperture 
telescope. Some of the fainter variables display rather large scatter 
due to observational noise. We left these stars in the sample, in 
order to test cases when the observational noise is more considerable, 
such as in the microlensing surveys of the Magellanic Clouds. A more 
thorough comparison of the set \#1 light curves will follow in the next 
subsection. Concerning set \# 2, we made a comparison with the periods 
published at the ASAS homepage. In general, we found that the periods 
agreed within $2\times10^{-5}$~d. In several instances our periods 
yielded better representations of the data (e.g., \object{V674~Cen}, 
\object{SS~For}, \object{VX~Scl}). 

The overall performance of the SNR-dependent Fourier fit is 
acceptable, but we see signatures of artifacts in some cases. 
These artifacts appear as unphysical waves and are related 
to as insufficient number of data points, sharp light curve variations 
or non-uniform coverage of the folded light curve. Some aspects of 
these sampling problems can be handled by the polynomial fitting 
method mentioned in JK96. In the present case this method performed 
slightly worse, i.e., yielded a few additional outliers in the 
comparison of the spectroscopic and Fourier metallicities. Therefore, 
we use the straightforward SNR-dependent Fourier fit, although in 
many cases the differences are small, mostly induced by noise.

%
%

\subsection{Comparison with the compilation of JK96}
Here, and in all subsequent comparisons of various datasets we 
follow a two-step procedure: (i) comparison of all available stars, 
disregarding any suspected source of defects (e.g., noise, systematics) 
or peculiarities (e.g., Blazhko effect); (ii) select outliers, check 
the agreement with the remaining sample and investigate the possible 
source(s) of discrepancy. This procedure likely leaves some stars in 
the sample that would otherwise be discarded. Nevertheless, we prefer 
this approach over the one followed by JK96, because the pre-selection 
procedure we applied there led to the omission of objects that 
otherwise fitted the calibration sample. In Sect.~3 we discuss other 
outlier selection strategies and the results obtained by their 
application.  

\begin{figure}[t]
   \centering
   \includegraphics[width=80mm]{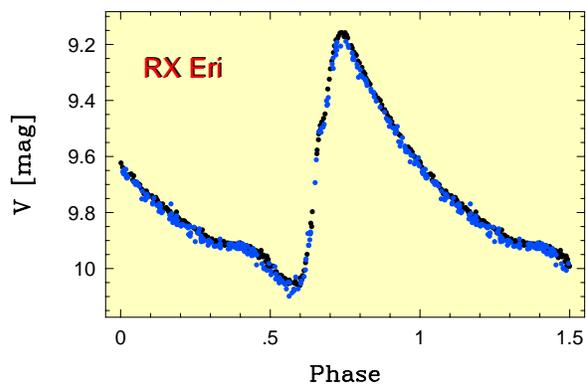}
      \caption{Comparison of the ASAS light curve of \object{RX~Eri} 
       (light dots) with that entering in the compilation of JK96 
       (black dots). The folding period of $0.58724599$~d was derived 
       from the combined dataset spanning 18.7~y. No adjustment of the 
       magnitudes was made.} 
         \label{fig4}
   \end{figure}

\begin{figure}[t]
   \centering
   \includegraphics[width=80mm]{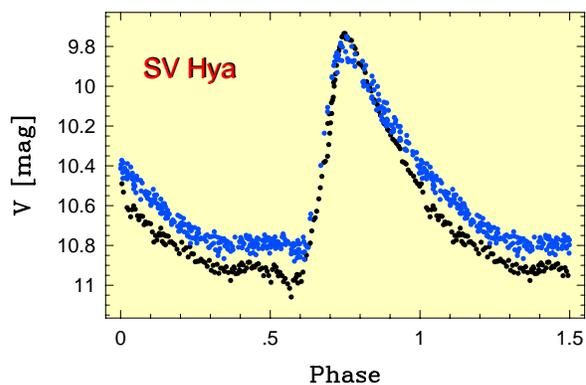}
      \caption{As in Fig.~\ref{fig4}, but for \object{SV~Hya}. Folding 
       period and total time span are 0.47854709~d and 33.3~y, respectively.} 
         \label{fig5}
   \end{figure}

Before making a more thorough comparison, we show two examples to 
display the range of differences between the two datasets. 
In Fig.~\ref{fig4} we display the type of agreement we have for 
about 75\% of the sample. Except for the adjustment of the period, 
no other transformations were made. The two light curves agree very 
well, with a small zero point offset of 0.016~mag. A different case 
is shown in Fig.~\ref{fig5}, where we observe changes both in the 
amplitude and in the average brightness. Although the variable was 
shown to be a Blazhko star from the ASAS data, we think that the 
difference can be attributed to an instrumental effect (e.g., blending, 
lack of color term in the magnitude transformation) rather than 
some, so-far unknown long-term effect of the Blazhko phenomenon. 
For example, the relatively large reddening of $E_{\rm B-V}=0.11$ 
(see L94), combined with the lack of a color term in the magnitude 
transformation may contribute to the discrepancy. In spite of the 
above differences, the phases are in good agreement, implying the 
stability of the [Fe/H] estimates even in these cases.
%
%
%
%
%
%
%
%
%
%
%
%
\begin{table*}[t]
\caption[ ]{Comparison of the light curve parameters}
\begin{flushleft}
\begin{tabular}{lclc}
\hline
Param. & $\Delta_1$ & Outliers & $\Delta_2$ \\
\hline
$P$   &  $(-3.0\pm 25.1)\times 10^{-6}$  
      &  \object{SV Hya}, 
         \object{IU Car}, 
	 \object{RX Cet}, 
	 \object{TV Leo}, 
         \object{RY Psc}, 
	 \object{AT Vir} 
      &  $(1.1\pm 3.9)\times 10^{-6}$ \\
$A_0$ &  $-0.016\pm0.044$  
      &  \object{AV Ser}, 
         \object{V445 Oph}, 
	 \object{SV Hya}, 
	 \object{RY Psc}, 
	 \object{FY Hya} 
      &  $\phantom{-}0.001\pm0.015$ \\
$A_1$ &  $-0.009\pm0.021$
      &  \object{RY Psc}, 
         \object{RX Cet}, 
	 \object{SV Hya}, 
	 \object{FY Hya}, 
	 \object{AV Ser}, 
	 \object{IU Car}, 
      & $-0.003\pm0.005$ \\ 
      &  & \object{V445 Oph}, 
           \object{RZ Cet}, 
	   \object{BB Pup} \\   
$A_2$ &  $-0.006\pm0.014$
      &  \object{RX Cet}, 
         \object{SV Hya}, 
	 \object{IU Car}, 
	 \object{AV Ser}
      &  $-0.001\pm0.007$ \\ 
$A_3$ &  $-0.003\pm0.012$
      &  \object{RX Cet} 
      &  $-0.001\pm0.008$ \\ 
$A_4$ &  $-0.002\pm0.012$
      &  \object{RX Cet}, 
         \object{RY Psc}, 
	 \object{IU Car}
      &  $-0.001\pm0.005$ \\ 
$\varphi_{21}$ &  $\phantom{-}0.010\pm0.065$
      &  \object{RZ Cet}, 
         \object{RX Cet}
      &  $\phantom{-}0.011\pm0.055$ \\ 
$\varphi_{31}$ &  $\phantom{-}0.020\pm0.114$
      &  \object{RX Cet}, 
         \object{IU Car}
      &  $\phantom{-}0.000\pm0.088$ \\ 
$\varphi_{41}$ &  $\phantom{-}0.050\pm0.132$
      &  \object{RX Cet}, 
         \object{TV Lib}, 
	 \object{DH Hya}, 
	 \object{FY Hya}, 
      &  $\phantom{-}0.010\pm0.083$ \\ 
$\rm [Fe/H]$ &  $\phantom{-}0.026\pm0.153$
      &  \object{RX Cet}, 
         \object{IU Car}
      &  $\phantom{-}0.001\pm0.119$ \\ 
\hline
\end{tabular}
\end{flushleft}
{\footnotesize\underline {Notes:} 
$\Delta_1 \equiv \langle p({\rm ASAS})-p({\rm JK96})\rangle $, 
where $p$ denotes the parameter entering in the first column, the 
$\langle ... \rangle$ symbol stands for simple average over the sample 
of the 31 stars of set \#1. The standard deviation of $\Delta_1$ 
is listed after the $\pm$ sign; 
$\Delta_2$ as above, but for the set obtained after $\sim 3\sigma$ 
outlier selection; 
outliers are listed in decreasing order of outlier status; 
$P$ stands for the period, $A_0$ for the average magnitude, 
$A_i$ and $\varphi_{i1}$ for the Fourier amplitudes and phases, 
respectively; 
units are in [d], [mag], [rad] and in [dex].                                               
}                                          
\end{table*}

In order to assess the overall agreement between the two datasets, 
we compare the period, the average brightness, the Fourier parameters 
(up to order 4) and [Fe/H] for set \#1. Table~3 shows the average 
differences and the standard deviations before and after outlier 
selection. We employed a $\sim 3\sigma$ criterion for outlier 
selection (here $\sigma$ stands for the standard deviation of the 
differences between the corresponding parameter values derived from 
the ASAS and JK96 sets). 

We see that the ASAS amplitudes are (on average) slightly smaller 
than those of JK96. An opposite trend can be observed for the phases. 
Although the effect is present in all amplitudes and phases, it is 
near to the $1\sigma$ error limits of these deviations, as can be 
seen from the standard deviations and the sample size. The standard 
deviations of the differences ($p({\rm ASAS})-p({\rm JK96})$, where 
$p$ denotes one of the parameters tested) are fairly small. In terms 
of the total ranges of the corresponding parameters, they correspond 
to $3$--$10$\% relative scatters. 

Pojmanski (2002) also performed a consistency test of the photometric 
zero points between the ASAS and Hipparcos databases. He concluded 
that the difference depended on the position of the object on the 
CCD frame, and in the best case the standard deviation of the 
differences was $0.018$~mag in the range of $8$--$12$~mag. Our 
comparison on the limited sample of set \#1 supports this result.   

From the point of view of the consistency of the [Fe/H] values, after 
selecting the two outliers, we have an agreement with a standard 
deviation of $0.119$~dex between the individual estimates. If we 
extrapolate this standard deviation as a general error of the currently 
available photometry and we take into consideration that the total 
standard deviation in the JK96 calibration was only $0.134$~dex, we 
end up with an unrealisticly small error of $0.062$~dex for the 
spectroscopic data (assuming a quadratic relation for the errors). 
Although the $0.119$~dex standard deviation can be decreased down 
to $0.093$~dex by omitting several other, less significant 
outliers (\object{RY~Psc}, \object{TV~Leo}, \object{TY~Pav}, 
\object{U~Lep} and \object{SX~For}), we think that except if we have 
long-term accurate photometry, in general we may expect an inaccuracy 
in the derived Fourier parameters that leads to an error near 
$0.1$~dex in the predicted iron abundance.    

It is difficult to pinpoint the leading source of the lack of 
higher-level of compatibility between the two datasets. In the case 
of the older photoelectric data, entering in the compilation of JK96, 
we usually have short tracks of data sometimes spanning several 
years. The tracks are, in general, from different sources. 
Construction of smooth folded light curves from these segments is 
often ambiguous because of the non-uniqueness of the combination 
of the data separated by long gaps. The advantage of the ASAS data 
over these older observations is that they are compact, and that 
the number of data points per star is relatively large. The 
disadvantage is the relatively high noise level and the lack of a more 
accurately defined magnitude system. This latter point was partially 
tested by checking the interrelations among the Fourier parameters 
(see JK96 and Kov\'acs \& Kanbur 1998). We found a very good overall 
agreement with the earlier derived relations, which is an indication 
that the ASAS light curves for RR~Lyrae stars must be fairly close 
to the standard $V$ system.   

In discussing the individual outliers listed in Table~3, four stars, 
\object{RX~Cet}, \object{RZ~Cet}, \object{SV~Hya} and \object{RY~Psc} 
turned out to be Blazhko stars (see Table~1). These variables often 
appear among the outliers in Table~3. In principle, if they had been 
observed frequently enough in both datasets, we would expect to 
recover the same average light curves. Unfortunately, as already 
mentioned above, the older data are rather sporadic and the ASAS 
data are somewhat noisy. These result in different Fourier 
decompositions in the two datasets. Variable \object{FY~Hya} is 
reasonably noisy in both datasets. For \object{V445~Oph} we suspect 
that the high reddening of $E_{\rm B-V}=0.19$ (see L94) might be 
instrumental in the observed differences for $A_0$ and $A_1$. 
Variable \object{AV~Ser} is reasonably noisy in the ASAS set with 
a relatively modest number of data points. The extremely short 
periodic star \object{TV~Lib} appears as an outlier only in 
$\varphi_{41}$. We do not know the cause of this, but this special 
object should be further analyzed, because if it is indeed an RRab 
star, it is very unique (we do not know any other variable classified 
as RRab-type with such a short period). 

The period analysis of the ASAS data on \object{IU~Car}, \object{TV~Leo} 
and \object{AT~Vir} yielded significantly different periods from the 
ones used by JK96 (the same is true for three Blazhko stars, but they 
have already been discussed). Among these period changing stars only 
\object{IU~Car} shows different light curves in the two datasets. 
A closer inspection of the JK96 data shows that only the data between 
HJD~2441807 and 2441695 are discrepant, the rest fit the ASAS data. 
It can also be seen that a shift of $\sim0.16$~d in the above interval 
cures the problem. Because of the large gaps in the sampling of 
\object{IU~Car} in the JK96 set, a separate frequency analysis of the 
light curve of this star in this dataset leads to the selection of an 
alias period. This solution no longer works when one tries to combine 
the data with the ASAS set. We note that \object{IU~Car} was an 
`unexplained' outlier in the [Fe/H] fit of JK96. With the correct 
period its outlier status disappears. 
       
Variables \object{DH~Hya} and \object{BB~Pup} are marginal outliers. 
Their status can be changed by a slight modification of the Fourier 
fitting algorithm, i.e., by using the polynomial fitting method 
mentioned in JK96.

%
%

\section{Predicted versus observed abundances}
Following the same methodology as in the previous section, here 
we make a comparison between the spectroscopic and the Fourier 
metallicities. We discuss the outlying objects, where the outlier 
status is defined by the arbitrary limit of $0.4$~dex between the 
difference of the spectroscopic and Fourier metallicities. We note 
that this limit corresponds to $\sim 3\sigma$ deviation according 
to the fit of JK96.

%
%

\subsection{Set \#1}
\begin{figure}[t]
   \centering
   \includegraphics[width=80mm]{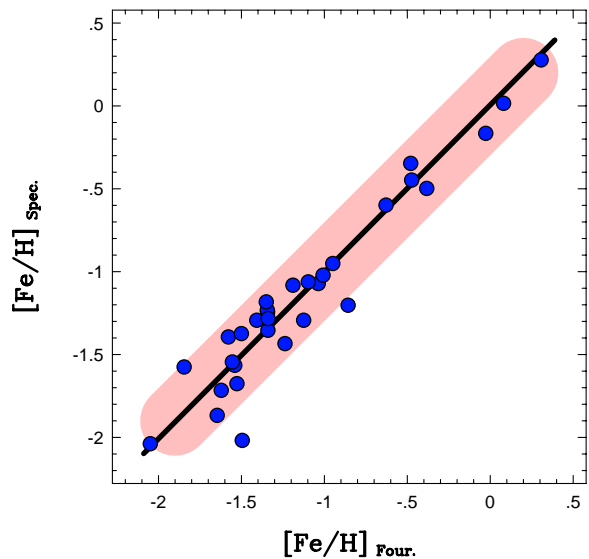}
      \caption{Spectroscopic versus Fourier [Fe/H] computed on 31 stars 
       of the ASAS database (set \#1, see Table~1). Shaded area shows 
       the $\pm 0.3$~dex boundaries around the 
       ${\rm [Fe/H]}_{\rm Spec.}={\rm [Fe/H]}_{\rm Four.}$ line.}  
         \label{fig6}
   \end{figure}

In Sect.~2 we compared the Fourier parameters and the predicted 
[Fe/H] of this set with those of JK96. Here we make a comparison 
only with the spectroscopic [Fe/H]. In Fig.~\ref{fig6} we show 
the correlation between the Fourier and spectroscopic [Fe/H]. 
With the criterion mentioned above, we found a single outlier. 
For compatibility with the discussion in the next subsection, we 
summarize the relevant properties of this star in Table~4. The NL 
classification of the light curves is somewhat subjective but 
reasonable support was lent by the few tests we made on the formal 
error of $\varphi_{31}$. The spectroscopic abundance was declared 
inaccurate (IS) whenever the number of measurements was lower than 
three or the values measured had standard deviation greater than 
0.20. We note that in the database of L94 the number of measurements 
for the individual objects rarely exceeds five.     

Considering the individual objects, we note that the light curve of 
\object{TY~Pav} is fairly noisy and the spectroscopic abundance is 
based only on two measurements. In the sequence of the degree of 
discrepancy, the next star is \object{RX~Cet}, with 
$\Delta{\rm [Fe/H]}=-0.34$~dex. This variable shows a strong Blazhko 
effect and the number of data points seems to be insufficient to  
obtain a reliable mean light curve that might yield a better fit to 
the observed abundance (see Alcock et al. 2003 for the similarity of 
the average light curves of the BL stars to those of the mono-periodic 
ones). \object{IU~Car} is no longer an outlier. With the ASAS data, 
its predicted abundance differs slightly less than 0.3~dex from the 
spectroscopic one. For the full sample of 31 stars 
$\sigma(\Delta{\rm [Fe/H]})=0.16$~dex, whereas with the omission of 
the single outlier we get $0.13$~dex, i.e., the same level of agreement 
as that of JK96.   
%
%
%
%
%
%
%
\begin{table}[t]                                                      
\caption[ ]{Outliers in [Fe/H] from set \#1}                                                    
\begin{flushleft}                                                     
\begin{tabular}{rrl}                                                  
\hline                                                                
Variable & $\Delta$[Fe/H] & Rem\\                                     
\hline
\object{TY Pav}     & $-0.51$ & NL, IS\\
\hline                                                                
\end{tabular}                                                         
\end{flushleft}                                                       
{\footnotesize\underline {Notes:}                                     
$\Delta$[Fe/H]=[Fe/H]$_{\rm Spec}-$[Fe/H]$_{\rm Four}$;               
BL : Blazhko variable;                                                
NL : Noisy light curve;                                               
IS : Inaccurate [Fe/H]$_{\rm Spec}$; 
See text for more comments}                                                      
\end{table}                                                           
%

%
%

\subsection{Set \#2}
This set is completely different from the calibration set used by JK96. 
Therefore, the test presented here constitutes the type of test one is 
supposed to perform when checking the predictive power of the Fourier 
method. 
%
%
%
   \begin{figure}[t]
   \centering
   \includegraphics[width=80mm]{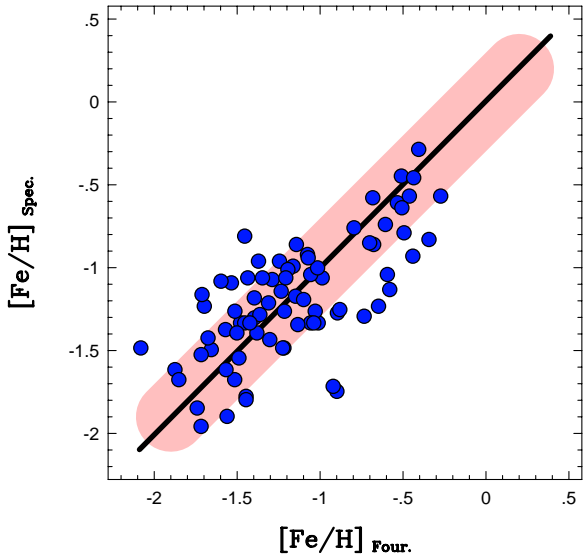}
      \caption{Spectroscopic versus Fourier [Fe/H] computed on 79 stars 
       of the ASAS database (set \#2, see Table~2). Notation is the same 
       as in Fig.~\ref{fig6}}  
         \label{fig7}
   \end{figure}

In a similar presentation as above, Fig.~\ref{fig7} shows the correlation,  
and Table~5 lists the outliers. We see that there are more outliers 
than in the case of set \#1. This increase is attributed to the lower 
brightness of the variables in set \#2. Because they are fainter, they 
are less accurately measured both photometrically and spectroscopically. 
At $V\approx 13$~mag, we are near the faint magnitude limit of the ASAS 
survey. 

%
%
%
\begin{table}[t]                                                      
\caption[ ]{Outliers in [Fe/H] from set \#2}                                                    
\begin{flushleft}                                                     
\begin{tabular}{rrl}                                                  
\hline                                                                
Variable & $\Delta$[Fe/H] & Rem\\                                     
\hline                                                                
\object{CK Aps}     & $ 0.47$ & NL, BL, IS\\
\object{DD Aps}     & $ 0.64$ & NL\\
\object{EL Aps}     & $-0.56$ & IS\\
\object{ER Aps}     & $-0.55$ & NL, IS\\
\object{UU Cet}     & $-0.84$ & IS\\
\object{SX Dor}     & $-0.79$ & NL\\
\object{SS For}     & $ 0.44$ & BL, IS\\
\object{SZ Hya}     & $ 0.60$ & BL\\
\object{DG Hya}     & $ 0.55$ & $-$\\
\object{FX Hya}     & $-0.49$ & IS\\
\object{IV Hya}     & $ 0.52$ & NL, IS\\
\object{V1023 Oph}  & $ 0.41$ & NL\\
\object{TZ Phe}     & $-0.45$ & NL\\
\object{V765 Sco}   & $-0.49$ & IS\\
\object{AS Vir}     & $-0.58$ & NL\\
\hline                                                                
\end{tabular}                                                         
\end{flushleft}                                                       
{\footnotesize\underline {Notes:} 
Notation is the same as in Table~4.}                                                      
\end{table}                                                        

We see that except for \object{DG~Hya}, {\em all} outliers can be 
explained either by high noise in the light curve, the presence of 
the Blazhko effect or by possible large errors in the abundance. 
Extreme examples for cases when the noisy light curves are the most 
probable causes of the discrepancies are \object{DD~Aps} and 
\object{SX~Dor}. There are several cases when the spectroscopy can 
most probably be blamed for the discrepancy. For example, the [Fe/H] 
values employed in this comparison are based on single measurements 
for \object{ER~Aps}. (Two other stars, \object{LU~Aps} and 
\object{AC~Eri}, with lower degree of discrepancy, also have this 
problem). 

The star \object{DG~Hya} stands alone with its significant outlier
status, not obviously related to its light curve or measured abundance. 
Indeed, from the Fourier fit and by the application of the error formula 
of JK96, we get $\sigma({\rm [Fe/H]}_{\rm Four})=0.10$. From the standard 
deviation and the number of measurements given by L94, we get 
$\sigma(\langle{\rm [Fe/H]}_{\rm Spec}\rangle)=0.07$. Assuming that 
these errors are independent, they yield a combined $1\sigma$ error 
of $0.12$~dex. This makes the observed discrepancy significant above 
the $4\sigma$ level. We note that for \object{DG~Hya} one may end up 
with a less discrepant Fourier abundance by using the polynomial 
fitting method mentioned Sect.~2.     

In the above outlier selection we relied on objects that correlate 
tightly and did not make a pre-selection based on the error estimate 
used above. As already mentioned, the reason for taking this approach 
is that objects with large errors may accidentally follow the 
correlation, thereby helping in the selection of the outliers. There 
are several stars (e.g., \object{UY~Aps}, \object{T~Men}, 
\object{RX~Men}, etc.) that do not enter in the list of outliers, 
although it is obvious that they have large errors in their Fourier 
parameters. 

The statistical error formula of JK96 quantifies the ambiguity due to 
light curve problems and can be used to pre-select candidates that 
are supposed to yield more accurate Fourier abundances. When applied 
to the present sample, we may end up with very limited samples by 
setting the error cutoff at values more preferable for accurate 
predictions. For example, with $\sigma({\rm [Fe/H]}_{\rm Four})$ 
cutoffs of $0.3$, $0.2$, $0.15$ and $0.1$, the sample size decreases 
to $67$, $57$, $48$ and $25$, respectively. Unfortunately, the 
spectroscopic abundances also have errors, therefore, we need to 
make further cuts in these already-limited samples if we want to 
consider all errors consistently. Without this latter cut, even 
the smallest sample of 25 stars in the above selection yields a 
large scatter of $0.22$~dex, indicating the presence of spectroscopic 
errors. 

With all $79$ stars left in the sample, we have 
$\sigma(\Delta{\rm [Fe/H]})=0.31$~dex. By leaving out the 15 outliers 
as defined at the beginning of this section and listed in Table~5, 
we get $0.20$~dex. This is in reasonable agreement with the expected 
scatter based on the calibration set of JK96 if we admit that the 
data of this separate sample are in general of lower quality than 
those of the calibrating set and that we left obviously bad quality 
light curves in the sample.    

Further trimming of the sample by applying cuts of $0.3$ and $0.2$~dex 
leads to sample sizes and dispersions of $55$, $39$ and $0.17$, $0.12$, 
respectively. We can also repeat this selection procedure on the 
database obtained by the application of the polynomial Fourier fitting 
method of JK96. We end up with similar result, yielding, e.g., a 
dispersion of $0.16$~dex for the remaining $52$ stars at the cutoff 
value of $0.3$~dex.

%
%

\subsection{The combination of sets \#2 and JK96}
\begin{figure}[t]
   \centering
   \includegraphics[width=80mm]{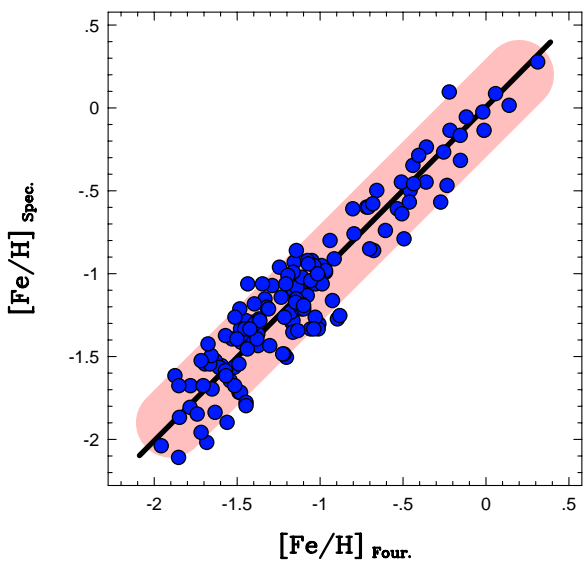}
      \caption{Spectroscopic versus Fourier [Fe/H] computed on the 
       outlier-free sample of 145 stars from the ASAS database and 
       from the compilation of JK96. See text for details on the 
       outlier selection. Notation is the same as in Fig.~\ref{fig6}}  
         \label{fig8}
   \end{figure}

To obtain a more comprehensive view of the fit to all available 
data that are free of outliers, we show the correlation for 145 
variables in Fig.~\ref{fig8}. This set contains the 64 stars remaining 
from the ASAS set \#2 and the 81 stars remaining from the basic set of 
JK96, after leaving out the three outliers \object{IU~Car}, \object{TZ~Aur}  
and \object{V341~Aql}. The standard deviation of the 145 stars is 
$0.17$~dex, not far from that of the calibrating set of JK96. 

In discussing the outliers, we recall that \object{IU~Car} has already 
been dealt with in Sect.~2.1. The outlier status of \object{IU~Car} 
disappeared by the use of ASAS data and the strongly discrepant 
abundance ($\Delta{\rm [Fe/H]}=0.61$) in the JK96 set is accounted 
for by a mysterious time shift in one of the data segments of that set. 
The variable \object{TZ~Aur} has a highly non-uniformly sampled light 
curve, with only 8(!) data points on the descending branch. Actually, 
with a straightforward Fourier method it cannot be fitted and one has 
to use a more sophisticated method, such as polynomial fitting 
preceding the Fourier decomposition, as mentioned in Sect.~2. 
Furthermore, there is only one measurement for [Fe/H] in L94 and the 
five measurements of S94 are rather noisy, yielding an error 
($1\sigma$ range) of $0.11$~dex in the published mean metallicity. 
\object{V341~Aql} has $\Delta{\rm [Fe/H]}=0.41$~dex, with an 
apparently well-established abundance, with two and ten measurements 
in the compilations of L94 and S94, respectively. The light curve has 
low scatter, spanning nearly four years with more than 500 data points. 
It seems to us that the outlier status of \object{V341~Aql} cannot be 
explained by observational noise.  

In summary, in the full sample of 163 stars consisting of the ASAS 
database and the compilation of JK96, we found only two stars, 
\object{DG~Hya} and \object{V341~Aql}, whose observed abundances cannot 
be predicted within 0.4~dex with the Fourier formula of JK96 and whose 
discrepant status is not easily explained by observational errors only. 
%
%
%

%
%
%

\section{Conclusions}
With the accumulating photometric data on RR~Lyrae stars, it becomes 
increasingly interesting to utilize methods that are capable of 
deriving physical parameters based on the morphology of the light 
curves. Of course, all methods should be thoroughly tested before they 
are applied on a target dataset. In this paper we performed such a 
test on the metallicity determination of the fundamental mode RR~Lyrae 
stars. 

The calibration is based on a sample of 81 Galactic field RRab stars 
by Jurcsik \& Kov\'acs (1996). The testing set comes from the ASAS 
database (Pojmanski 2002, 2003; Pojmanski \& Maciejewski 2004). Only 
those variables were selected from the database that had corresponding 
low-dispersion spectroscopic metallicities either in Layden (1994) or 
in Suntzeff et al. (1994). The final ASAS sample contains 110 stars, 
31 of them overlap with the compilation of Jurcsik \& Kov\'acs (1996). 
We did not apply any {\em quality control} in the selection procedure. 
We use all light curves irrespective of whether they have mono-periodic 
light curves of low- or high-noise or show the Blazhko effect. This 
approach is opposite to the one we followed in the construction of the 
calibration set. The purpose of the present approach is to make 
allowance for less favorable cases, when the high noise level or the 
short time span of the observations do not make a more critical selection 
possible. 

The result of the comparison shows that almost all objects deviating 
by more than $0.4$~dex from the spectroscopic values exhibit some 
(well-demonstrated) problems either in the light curve or in the 
spectroscopy. We have only two objects (\object{V341~Aql} and 
\object{DG~Hya}) whose discrepant positions cannot be explained by 
these simple causes. The outlier-free sample of 64 stars distinct 
from the calibrating sample reproduces the spectroscopic abundances 
with a standard deviation of $0.20$~dex. When this ASAS sample is 
combined with the original calibrating sample of 81 stars (also free 
of outliers), the standard deviation becomes $0.17$~dex, not far 
from the fitting accuracy of the calibrating set.   
   
Although the above result is satisfactory in terms of the dispersion 
between the computed and measured metallicities, it may be far from 
the accuracy we could reach with better metallicities and light curves. 
The quality of the light curves can be improved by employing larger 
aperture telescopes. With the current sophisticated data reduction 
methods (e.g., image subtraction, see Alard 2000) an accuracy of 
$0.01$--$0.02$~mag per data point could be routinely reached with 
a $\sim 20$~cm telescope for stars of $12$--$13$~mag. Long-term 
monitoring (from months to years) is necessary to obtain sufficient 
coverage for Blazhko variables. Based on the result of Alcock et al. 
(2003), it is possible that the average light curves of the Blazhko 
stars can be employed equally well in the empirical formulae calibrated 
on strictly periodic stars. This conjecture is partially supported by 
the present study, because several Blazhko stars did stay in the set 
after outlier selections. It is highly probable that the situation 
can be further improved by gathering more data points and covering 
the Blazhko cycle more fully. 

A real improvement may come from a renewed effort to gather more 
accurate spectroscopic data. Ambiguities due to pulsation phase 
dependence should be minimized by taking more observations per object. 
The fact that globular cluster giants are often measured with an 
accuracy higher than $0.1$~dex implies that high-dispersion data 
could reach similar accuracy once the phase dependence is appropriately 
taken care of. The goal of good phase coverage can, of course, 
be reached more easily by taking low-dispersion spectra and using 
a spectral index method. However, these spectral index methods should 
be calibrated directly by the updated accurate high-dispersion 
abundances mentioned above. 

To target individual variables with the Fourier method and aiming 
at accurate abundance estimates on specific objects, one has to 
be sure that the formula is applicable to any RRab star. Currently, 
with the two unexplained outliers, we have a chance of $\sim 1$\% of  
hitting such an object in a randomly selected target list. Although 
this is a relatively small chance, for a more consistent assessment 
of the outlier status of the individual objects, we would need better 
error estimates of the spectroscopic abundances. With the few data 
points per variable and the lack of more precise knowledge of the 
phase dependence of the [Fe/H] estimates, the required reliable error 
estimation is not possible. 

In principle, if the relation between [Fe/H] and the Fourier parameters 
indeed holds for the overwhelming majority of RRab stars, because of 
the general statistical property of a regression to noisy data, we 
expect more accurate [Fe/H] estimates from the Fourier method than from 
direct spectroscopic measurements. Therefore, it will become more 
interesting to address the physical origin of this relation. 
Unfortunately, the cause of this relation is unknown at the moment. 
On the empirical side, in a recent paper Sandage (2004) attempts to 
derive the $P$--$\varphi_{31}$--[Fe/H] relation on the basis of other 
empirical relations, exploring the correlations among the various 
light curve parameters, without giving any explanation for the existence 
of the formula for [Fe/H]$_{\rm Four}$. On the theoretical side, except 
for a note by Bono, Incerpi \& Marconi (1996) and for some partial 
results of Feuchtinger \& Dorfi (1997) and Dorfi \& Feuchtinger (1999) 
there are no works targeting the background of this relation. Although 
in the latter papers we find that the models show the predicted trends 
(i.e., increase of $\varphi_{31}$ with the increase of $P$, shifting 
the [Fe/H]=const ridge lines on the $P\rightarrow\varphi_{31}$ plots), 
no quantitative comparison is possible because of the limited scope of 
the theoretical surveys. This situation further emphasizes the need for 
accurate spectroscopic abundances that are capable of helping in the 
derivation of a tighter calibration formula, whereby the theoretical 
models can be confronted with more stringent empirical constraints. 

Until then, with the current fitting accuracy of $0.17$~dex of 
the Fourier formula, we conservatively estimate the {\em overall} 
prediction accuracy, i.e., the standard deviation of the predicted 
[Fe/H], to be $0.12$~dex (by sharing the total error budget equally 
between the Fourier parameters and the spectroscopic abundances). 
This accuracy makes the Fourier method a useful tool for the 
estimation of the abundances of various populations of RR~Lyrae stars 
but limits its applicability in individual cases, where higher accuracy 
might be demanded.

\begin{acknowledgements}
We thank Andrew Layden for the useful correspondence on the current 
status of the spectroscopic abundances. This research has made use of 
the SIMBAD database, operated at CDS, Strasbourg, France. The support 
of {\sc otka} grant {\sc t--038437} is acknowledged.
\end{acknowledgements}


\end{document}